\documentstyle [12pt,epsf,rotate] {article}

\newcommand{\nwc}{\newcommand}
\nwc{\cl}  {$\clubsuit$}

\nwc{\hyp} {\hyphenation} 
\nwc{\be}  {\begin{equation}}
\nwc{\ee}  {\end{equation}}
\nwc{\ba}  {\begin{array}}
\nwc{\ea}  {\end{array}}
\nwc{\bdm} {\begin{displaymath}}
\nwc{\edm} {\end{displaymath}}
\nwc{\bea} {\be\ba{rcl}}
\nwc{\eea} {\ea\ee}
\nwc{\ben} {\begin{eqnarray}}
\nwc{\een} {\end{eqnarray}}
\nwc{\bda} {\bdm\ba{lcl}}
\nwc{\eda} {\ea\edm}
\nwc{\bc}  {\begin{center}}
\nwc{\ec}  {\end{center}}
\nwc{\ds}  {\displaystyle}
\nwc{\bmat}{\left(\ba}
\nwc{\emat}{\ea\right)}
\nwc{\non} {\nonumber}
\nwc{\bib} {\bibitem}
\nwc{\lra} {\longrightarrow}
\nwc{\Llra}{\Longleftrightarrow}
\nwc{\ra}  {\rightarrow}
\nwc{\Ra}  {\Rightarrow}
\nwc{\lmt} {\longmapsto}
\nwc{\prl} {\partial}
\nwc{\iy}  {\infty}
\nwc{\ol}  {\overline}
\nwc{\hm}  {\hspace{3mm}}
\nwc{\lf}  {\left}
\nwc{\ri}  {\right}
\nwc{\lm}  {\limits}
\nwc{\lb}  {\lbrack}
\nwc{\rb}  {\rbrack}
\nwc{\ov}  {\over}
\nwc{\pri}  {\prime}
\nwc{\nnn} {\nonumber \vspace{.2cm} \\ }
\nwc{\Sc}  {{\cal S}}
\nwc{\Lc}  {{\cal L}}
\nwc{\Rc}  {{\cal R}}
\nwc{\Dc}  {{\cal D}}
\nwc{\Oc}  {{\cal O}}
\nwc{\Cc}  {{\cal C}}
\nwc{\Pc}  {{\cal P}}
\nwc{\Mc}  {{\cal M}}
\nwc{\Ec}  {{\cal E}}
\nwc{\Fc}  {{\cal F}}
\nwc{\Hc}  {{\cal H}}
\nwc{\Kc}  {{\cal K}}
\nwc{\Xc}  {{\cal X}}
\nwc{\Gc}  {{\cal G}}
\nwc{\Zc}  {{\cal Z}}
\nwc{\Nc}  {{\cal N}}
\nwc{\fca} {{\cal f}}
\nwc{\xc}  {{\cal x}}
\nwc{\Ac}  {{\cal A}}
\nwc{\Bc}  {{\cal B}}
\nwc{\Uc}  {{\cal U}}
\nwc{\Vc}  {{\cal V}}
\nwc{\Th} {\Theta}
\nwc{\th} {\theta}
\nwc{\vth} {\vartheta}
\nwc{\eps}{\epsilon}
\nwc{\si} {\sigma}
\nwc{\Gm} {\Gamma}
\nwc{\gm} {\gamma}
\nwc{\bt} {\beta}
\nwc{\La} {\Lambda}
\nwc{\la} {\lambda}
\nwc{\om} {\omega}
\nwc{\Om} {\Omega}
\nwc{\dt} {\delta}
\nwc{\Si} {\Sigma}
\nwc{\Dt} {\Delta}
\nwc{\al} {\alpha}
\nwc{\vp} {\varphi}
\nwc{\kp} {\kappa}
\def\tr{\mathop{\rm tr}}

\def\ltap{\raisebox{-.4ex}{\rlap{$\sim$}} \raisebox{.4ex}{$<$}}
\def\gtap{\raisebox{-.4ex}{\rlap{$\sim$}} \raisebox{.4ex}{$>$}}
\nwc{\Id}  {{\bf 1}}
\nwc{\diag} {{\rm diag}}
\nwc{\inv}  {{\rm inv}}
\nwc{\mod}  {{\rm mod}}
\nwc{\hal} {\frac{1}{2}}
\nwc{\tpi}  {2\pi i}

\def\npb#1{Nucl.\ Phys.\ {\bf B#1}}

\def\plb#1{Phys.\ Lett.\ {\bf B#1}}

\def\pra#1{Phys.\ Rev.\ {\bf A#1}}

\def\prd#1{Phys.\ Rev.\ {\bf D#1}}
\def\prle#1{Phys.\ Rev.\ Lett.\ {\bf #1}}

\newsavebox{\nnin} \sbox{\nnin}{$\hspace{1mm}\in\kern -.8em /
                   \hspace{1mm}$}

\newcommand{\sub}{\subset}
\newsavebox{\nnsub} \sbox{\nnsub}{$\hspace{1mm}\sub\kern -.9em /
            \hspace{1mm}$}

\def\KK{{\rm I\kern -.2em  K}}
\def\NN{{\rm I\kern -.16em N}}
\def\RR{{\rm I\kern -.2em  R}}
\def\ZZ{Z \kern -.43em Z}
\def\QQ{{\rm \kern .25em
             \vrule height1.4ex depth-.12ex width.06em\kern-.31em Q}}
\def\CC{{\rm \kern .25em
             \vrule height1.4ex depth-.12ex width.06em\kern-.31em C}}
\def\ZZZ{Z\kern -0.31em Z}

\nwc{\olnu}  {\ol{\nu}}
\nwc{\olla}  {\ol{\la}}
\nwc{\olm}   {\ol{m}}
\nwc{\olmu}  {\ol{\mu}}
\nwc{\olh}   {\ol{h}}
\nwc{\olpsi} {\ol{\psi}}
\nwc{\olsi}  {\ol{\sigma}}
\nwc{\olgm}  {\ol{\gm}}
\nwc{\prlt}  {\frac{\prl}{\prl t}}
\nwc{\ttau}  {\tilde{\tau}}
\nwc{\trho}  {\tilde{\rho}}
\nwc{\tP}    {\tilde{P}}
\nwc{\tU}    {\tilde{U}}
\nwc{\teps}  {\tilde{\eps}}
\nwc{\tla}   {\tilde{\la}}
\nwc{\tit}    {\tilde{t}}
\nwc{\iddq}  {\int\frac{d^dq}{(2\pi)^d}}
\nwc{\prpr}  {\prime\prime}
\nwc{\rN}    {\left(\frac{\rho}{N}\right)}
\nwc{\rNt}    {\left(\frac{\rho}{N}\right)^{\frac{N-2}{2}}}
\nwc{\rnN}   {\left(\frac{\rho_0}{N}\right)}
\nwc{\rnNt}    {\left(\frac{\rho_0}{N}\right)^{\frac{N-2}{2}}}
\nwc{\rnNf}    {\left(\frac{\rho_0}{N}\right)^{\frac{N-4}{2}}}
\nwc{\rNs}    {\left(\frac{\rho_0}{N}\right)^{\frac{N-6}{2}}}
\nwc{\kNt}    {\left(\frac{\kappa}{N}\right)^{\frac{N-2}{2}}}
\nwc{\kNf}    {\left(\frac{\kappa}{N}\right)^{\frac{N-4}{2}}}
\nwc{\kNs}    {\left(\frac{\kappa}{N}\right)^{\frac{N-6}{2}}}

\begin{document}
\begin{titlepage}

\title{ Coarse Graining and First Order Phase Transitions\\ 
\ \\ }

\author{{\sc J.\ Berges$^a$\thanks{Email:
J.Berges@thphys.uni-heidelberg.de}},\ \ \ \
{\sc N.\ Tetradis$^b$\thanks{Email: Tetradis@mail.cern.ch}} 
\\ \\ and \\ \\
{\sc C.\ Wetterich$^a$\thanks{Email: 
C.Wetterich@thphys.uni-heidelberg.de}}
\\ \\ \\
{\em $^a$Institut f\"ur Theoretische Physik, 
Universit\"at Heidelberg,} \\
{\em Philosophenweg 16, 69120 Heidelberg, Germany}\\ \\
{\em $^b$CERN, Theory Division,}\\
{\em CH-1211, Geneva 23, Switzerland}}

\date{}
\maketitle

\begin{picture}(5,2.5)(-350,-450)
\put(10,-70){CERN--TH--96--289}
\put(10,-85){HD--THEP--96--042}
\end{picture}

\thispagestyle{empty}

\begin{abstract}

We discuss the dependence of the coarse grained free energy and the 
classical interface tension on the coarse
graining scale $k$. A stable range appears only if the 
renormalized dimensionless couplings at the critical temperature
are small. This gives a quantitative criterion for the validity
of computations within Langer's theory of spontaneous bubble nucleation.

\end{abstract}

\end{titlepage}

The discussion of the dynamics
of a first order phase transition \cite{langer}
usually relies on the study
of a non-convex potential or free energy. 
The decay of unstable minima 
is associated either with tunneling fluctuations through 
barriers in the potential \cite{callan}, or, at non-zero temperature, 
with thermal fluctuations above them
\cite{bubbles}. However, the 
effective potential \cite{effpot}, which seems 
at first sight a natural tool for 
such studies, 
is expected to be a convex quantity with no
barrier. The resolution of this paradox lies in the 
realization that the effective potential is convex because
the tunneling or thermal fluctuations are incorporated in it.
These fluctuations are associated with low frequency modes, while the
non-convex part of the potential is related to the 
classical potential and
the integration of high frequency modes. 
A natural approach to the study of first order phase transitions
separates the problem in two parts. First, the
high frequency modes are integrated out, with the possible
generation of new minima through radiative symmetry breaking \cite{colwein}. 
Subsequently, the decay of unstable minima is discussed with
semiclassical techniques \cite{callan,bubbles}, 
in the non-convex potential that has 
resulted from the first step. 
This leads us to the notion of the coarse grained free energy, which
is fundamental in statistical physics. Every physical system
has a characteristic length scale associated with it. The dynamics
of smaller length scales is integrated out, and
is incorporated in the parameters of the 
free energy one uses for the study of the behavior at larger length scales.

The notion of coarse graining is 
absent in the perturbative approach to the calculation of the
effective potential \cite{colwein}. This is the main reason 
for the non-convergence of the perturbative series near the 
maxima of the classical potential, and the appearance of 
imaginary parts in the perturbative effective potential. 
Despite attempts to give a physical interpretation to these imaginary
parts \cite{complex}, a satisfactory discussion of tunneling must
incorporate the notion of coarse graining.
The Wilson approach to the renormalization group 
provides the appropriate framework \cite{wilson}. 
We employ here the method of the 
effective average action $\Gamma_k$ \cite{averact}, which  
results from the integration of fluctuations with characteristic
momenta larger than a given scale $k$. 
The dependence of $\Gamma_k$ on $k$ is described
by an exact renormalization group equation\footnote{
For related work see \cite{wilson,erg,gauge2}.} 
\cite{averact,gauge}.
For large values of $k$ (of the order of the ultraviolet cutoff 
$\La$ of the theory) 
the effective average action is equal to the classical action 
(no fluctuations are integrated out), while for $k \rightarrow 0$
it becomes the standard effective action (all fluctuations are 
integrated out). 
For non-zero $k$ the effective average
action has the properties of a coarse grained free energy.  
Its non-derivative part (the effective average potential $U_k$) is 
not necessarily convex. The coarse graining scale 
can be identified with $k$.

In this letter we provide an explicit 
demonstration of how such a potential can 
be obtained starting from the 
microscopic or classical action of a field theory.  
We investigate the dependence of the effective average
potential $U_k$ and the `classical'
surface tension $\sigma_k$
on the coarse graining 
scale $k$ with special emphasis on the question
of the validity of Langer's 
theory of bubble formation. We study the first order 
phase transitions for the Abelian Higgs model
and for a scalar matrix model 
in three dimensions. 
An application of the formalism
to the case of the high temperature phase transitions for the
Abelian and $SU(2)$ Higgs models is given in ref.\ \cite{ew}
and a discussion of the 
first order phase transition in matrix models can be found 
in ref.\ \cite{matrix}.

Near a phase transition, 
the three dimensional Abelian Higgs model describes the behavior of 
ordinary superconductors \cite{supercond}. It can also be viewed as the 
effective theory resulting from the non-zero temperature four dimensional
model near the critical temperature. 
The dependence of the effective average potential $U_k(\rho)$ 
and the running renormalized gauge coupling $e_R(k)$ on the 
coarse graining scale $k$ is governed by the evolution equations 
\cite{gauge}
\bea
\ds{\frac{\partial U_k(\rho)}{\partial t}} &=& 
\ds{\int \frac{d^3q}{2(2 \pi)^3} \frac{\partial P_k}{\partial t}
\left( 
\frac{1}{P_k(q) + U_k'(\rho) + 2 U_k''(\rho) \rho} \right.} \nnn
&&\ds{\left. + \frac{1}{P_k(q) + U_k'(\rho)} 
\rule{0mm}{5mm} 
+ \frac{2}{P_k(q) + 2 e^2_R(k) \rho}
\rule{0mm}{5mm}  \right)},\nnn
\ds{\frac{d e^2_R(k)}{dt}}&=&  
\ds{\frac{0.84}{6 \pi^2} \frac{e^4_R(k)}{k}}
\label{one} 
\eea
where
$t = \ln (k/\La)$ (with $\La$ 
the ultraviolet cutoff of the theory) and 
$\rho=|\phi|^2/2$ (with $\phi$ the complex order parameter). 
Primes denote derivatives with respect to $\rho$.
The three terms on the r.h.s.\ correspond to the contributions of
the radial and Goldstone scalar modes and the gauge field.
The inverse propagator\be
P_k(q) = \frac{q^2}{1-
\exp \left( -{q^2}/{k^2} \right)}
\label{two} \ee
provides for an infrared cutoff which acts like 
a mass term $\sim k^2$ for the modes with $q^2 \ll k^2$, while
it leaves unaffected the modes with $q^2 \gg k^2$. 
The momentum integral on the r.h.s.\ of eq.\ (\ref{one})
can be written in terms of dimensionless functions
$l^3_0(w)$, whose arguments are given by the rescaled
mass terms $(U_k'(\rho) + 2 U_k''(\rho) \rho)/k^2$,
$U_k'(\rho)/k^2$ and $2 e^2_R(k) \rho/k^2$. 
These functions fall off for large values of $w$, 
following a power law. As 
a result they introduce threshold behavior, 
which leads to the decoupling of 
massive modes from the evolution equations
\cite{averact,convex,indices}.
The derivation of eq.\ 
(\ref{one}) under some approximations,
starting from the
exact renormalization group equation for the effective average action,
is given in ref. \cite{gauge,supercond,ew}. 
The approximations concern the 
omission of the anomalous dimension of the scalar field, 
the effective field dependence of the gauge coupling and
the higher derivative terms in the action.
The evolution starts at $k=\La$, where the effective average potential
is equal to the microscopic or classical one 
$U_{\La}(\rho)=\hal \bar{\la}_{\La}(\rho-\rho_{0\La})^2$ 
and the running gauge 
coupling is equal to the bare coupling $\bar{e}_{\La}$. 
In the opposite
limit $k \rightarrow 0$, $U_k(\rho)$ becomes equal to 
the (convex) effective potential
$U(\rho)=U_0(\rho)$ and the gauge coupling assumes its 
renormalized value $e_R=e_R(k=0)$. 
Two algorithms for the numerical integration of eq.\ (\ref{one})
have been presented in detail \cite{num}.
The phase 
transition is approached by fixing $\bar{\la}_{\La}$ 
and $\bar{e}_{\La}$ and 
tuning $\rho_{0 \La}$.
The system exhibits a second order phase transition
for $\bar{e}_{\La}=0$ which corresponds to the Wilson-Fisher
fixed point of the $O(2)$ symmetric Heisenberg model.
For large enough $\bar{e}^2_{\La}/\bar{\la}_{\La}$ 
the phase transition is first order \cite{hal,supercond,ew}.

In fig.\ 1 we display the solution of eq.\ (\ref{one}) for 
$\bar{\la}_{\La}=0.01 \La$, $\bar{e}^2_{\La}=0.1 \La$ and 
$\rho_{0 \La} \simeq 0.867 \La$. All the quantities in the figures 
are expressed in units of the ultraviolet cutoff $\La$.
Initially the potential has only one minimum 
away from the origin, which evolves
proportionally to the coarse graining scale $k$. 
At some point a new minimum appears at the origin. 
It is induced by 
the integration of fluctuations, through the
generalization of the Coleman-Weinberg mechanism.
The evolution slows down at the later stages, and 
for $k/\La$ around 0.02 the 
potential converges towards a stable non-convex profile with two
minima of equal depth. Around the minima
the scale $k$ becomes smaller
than the mass of the various massive modes,
and this induces their decoupling. 
We have stopped the evolution at a non-zero $k_{f}$, for which the
shape of the potential near the minima is stable. The presence of the
non-convex part is explained by this non-zero value of $k$.
We have not yet integrated out all the fluctuations, 
which should render the effective potential convex. More specifically,
the fluctuations which interpolate 
between the two minima of fig.\ 1 
are not included effectively
in the non-convex potential. They are the ones
that trigger the tunneling and drive the first order 
phase transition. If we continue the evolution all the way
to $k=0$, these interpolating configurations will be  
gradually integrated out.
As a result, the height of the barrier will start 
getting smaller, until
the region of the potential between the two minima becomes flat
\cite{convex}.
The evolution of the characteristics of the potential is depicted in
fig.\ 2. We plot the location of the minimum away 
from the origin $\rho_{min}$,
the value of the potential at the minimum
$(U_k)_{min}$, the location of the maximum
$\rho_{max}$, the value of the potential 
$(U_k)_{max}$ and the curvature 
$(d^2U_k/d \phi^2)_{max}=U'_k(\rho_{max})
+2  \rho_{max} U''_k(\rho_{max})$ at the maximum.
We observe that these parameters have almost constant values in the
region $k/\La \simeq 0.02-0.03$.

In fig.\ 3 and 4
we present the effective average potential 
and its characteristics for parameters corresponding to a
weaker first order transition. For   
$\bar{\la}_{\La}=0.1 \La$, $\bar{e}^2_{\La}=0.1 \La$ and 
$\rho_{0 \La} \simeq 0.171 \La$
the discontinuity in the scalar field expectation value
is about nine times smaller than for fig.\ 1. 
The most important difference 
is that the potential never 
becomes relatively stable for a range of $k$. 
During the later stages of the evolution, its outer 
part (for scalar field values
larger than the location of the minimum) starts approaching a 
stable profile, due to the decoupling of the massive modes 
in this region.
However, in the same range of $k$ the non-convex part 
starts already becoming flatter, as configurations 
interpolating between the
two minima are being integrated out. 
The negative curvature at the top of the barrier is expected 
\cite{convex} to
behave $\sim -k^2$ during this stage. This has been
verified explicitly through the analytical integration of
the evolution equation for the $O(N)$ symmetric scalar
theory in the large $N$ limit \cite{analy}. We clearly
observe the onset of this behavior of $d^2U/d \phi^2$
in fig.\ 4 in a range of $k$ before $\rho_{min}$ settles.
Also the maximum of $U_k$ decreases before the minimum 
settles. In this case it is far from obvious which
coarse graining scale $k$ should be chosen for a definition 
of important nucleation characteristics such as the interface 
tension.

The behavior of the coarse grained effective potential for a
fluctuation induced first order phase transition, as 
considered above, 
is not particular to the Abelian Higgs model. 
For example, our discussion can be extended with minor 
modifications to the electroweak phase transition for the 
range of Higgs-scalar masses where it is first order
\cite{ew}.
It also
can be observed in pure
scalar theories. One may consider models with $U(N) \times U(N)$ 
symmetry with
a scalar field in the $(\bar{N},N)$ representation,
described by a complex $N \times N$ matrix
$\phi$ \cite{matrix}. The cases $N=2$, $3$ have an 
interesting relation to high
temperature strong interaction physics 
and chiral symmetry breaking \cite{PRW} 
and non-perturbative flow equations have been studied 
in this context \cite{Ju95-7}.
We will concentrate here on $N=2$. The most
general effective average potential $U_k(\rho,\tau)$ can then
be expressed as a function of only two invariants, namely
\be
 \rho = \ds{\tr\left(\phi^\dagger \phi \right) },\quad
 \tau = \ds{2 \tr\left(\phi^\dagger \phi - \frac{1}{2} \rho \right)^2}.
 \label{Invariants}
\ee
The microscopic or classical potential 
$U_{\La}$ for these models can be
characterized by two quartic couplings $\bar{\la}_{1\La}$,
$\bar{\la}_{2\La}$ and a mass term $(\bar{\mu}_{\La}^2 > 0)$,
\be
U_{\La}(\rho,\tau)=-\bar{\mu}_{\La}^2 \rho + \hal \bar{\la}_{1\La}
\rho^2 +\frac{1}{4} \bar{\la}_{2\La} \tau \, .\quad
\label{uinitial}
\ee
In the limit
$\bar{\la}_{1\La}\to \infty$, $\bar{\la}_{2\La}\to \infty$
this also covers the model of unitary matrices.
One 
observes two symmetry breaking patterns for
$\bar{\la}_{2\La}>0$ and $\bar{\la}_{2\La}<0$
respectively. The case $\bar{\la}_{2\La}=0$
denotes the boundary between the two phases.
In this special case the theory
exhibits an enhanced $O(8)$ symmetry and one finds a second order
phase transition.  
For the symmetry breaking pattern
$U(2) \times U(2) \to U(2)$ ($\bar{\la}_{2 \La}>0$)
the phase transition is always first order. 
In this case the relevant 
information for the phase transition is contained in 
$U_k(\rho) \equiv U_k(\rho,\tau=0)$.
The discussion of the dependence of the 
effective average potential $U_k(\rho)$ on the coarse 
graining scale $k$
can be presented along the same lines as for the Abelian Higgs 
model and the relevant flow equations can be found in ref.\
\cite{matrix}. 
Here the second quartic coupling $\bar{\la}_{2\La}$
for the scalar model
plays the role of the gauge coupling $\bar{e}^2_{\La}$
in the Abelian Higgs model.
In addition to the above treatment of the Abelian Higgs model,
the employed approximation for the scalar model
takes into account a $k$-dependent wave function
renormalization constant $Z_k$ for the fields and
the effective field dependence of the second quartic coupling.

In the following we will use the scalar 
matrix model to establish a 
quantitative criterion for the validity of the standard 
treatment of spontaneous bubble nucleation as 
described by Langer's 
theory \cite{langer}.  
Langer's approach relies, on the one hand, on
the definition of a suitable
coarse grained free energy $\Gm_k$ with a coarse graining
scale $k$ and, on the other hand, on a saddle point
approximation for the treatment of fluctuations around
the `critical bubble'.
The problem is therefore separated in two parts:
The first part concerns the treatment of fluctuations with 
momenta $q^2 \, \gtap \, k^2$ which are included in the
coarse grained free energy. The second part deals with an 
estimate of fluctuations around the bubble,
for which only fluctuations with momenta smaller than
$k$ must be considered.
To be explicit we consider a spherical bubble
where the bubble wall with thickness $\Dt$
is thin as compared to the bubble radius $R$, i.e.
$\Dt \ll R$.
In this thin wall approximation
the bubble nucleation
rate $\bar{\Gm}$, which describes the probability
per unit volume per unit time for the transition
to the new vacuum, 
can be written in the form \cite{bubbles,callan}
\be
\bar{\Gm} = A_k \, \exp \left(-\frac{16 \pi}{3}\frac{\sigma_k^3}
{\eps^2}\right)
\label{nc}.
\ee 
Here the `classical'
surface tension $\sigma_k$ is given in our conventions by
\be
\sigma_k=2\int\limits^{\bar{\vp}}_0 d\vp
\sqrt{2 Z_k U_k(\vp)}
\label{sur}
\ee
where $\vp=(\rho/2)^{1/2}$ and 
$\bar{\vp}$ denotes the second zero of $U_k(\vp)$
near the outer
minimum at $\vp_{min}$. For
the difference in the free energy density $\eps$ 
we include fluctuations with arbitrarily
small momenta,
\be
\eps=\lim_{k \to 0}(U_k(0)-U_k(\vp_{min})).
\label{vo}
\ee
In contrast, the long wavelength
contributions to the true surface tension\footnote{
The true surface tension is a `measurable quantity'.
It is independent of $k$ and all fluctuations 
must be included. It therefore differs, in general,
from $\si_k$ which includes only part of the fluctuations.} 
are effectively cut off by the 
characteristic length scale of the bubble surface.
For the free energy of the critical bubble
the modes with $q^2 \, \gtap \, k^2$ 
are incorporated in $\Gm^{(0)}_k=16 \pi \sigma_k^3/3 \eps^2$
(lowest order or classical
contribution)
and they influence $\si_k$. 
The modes with $q^2 \, \ltap \, k^2$ contribute to 
the `fluctuation determinant'
$A_k$, which also contains dynamical factors. 
Here $A_k$ depends on $k$ 
through the effective ultraviolet
cutoff for these fluctuations, which is present
since fluctuations with momenta larger than $k$
are already included in 
$\Gm^{(0)}_k$. A more general discussion which does not rely
on the thin wall approximation or a saddle point
approximation is given in ref.\ \cite{matrix}.
Langer's formula for
bubble nucleation amounts essentially to a perturbative
one loop estimate of $A_k$.
It is clear that $k$ is only a technical construct
and for physical quantities 
like the bubble nucleation rate the $k$-dependence of
$\ln A_k$ and $\Gm_k^{(0)}$ must cancel.
A strong
dependence of $\si_k$ on the
coarse graining scale $k$ is only compatible with
a large contribution from the higher orders
of the saddle point expansion. The $k$-dependence of
$\si_k$ therefore
gives direct information about the 
convergence of the saddle point approximation
and the validity of 
Langer's formula. 

We consider in detail the dependence of
the surface tension $\sigma_k$ on the coarse
graining scale $k$ at the phase transition $(\eps=0)$
for three examples.
They are 
distinguished by different choices for the quartic couplings
$\bar{\la}_{1\La}$ and $\bar{\la}_{2\La}$ of the short
distance potential $U_{\La}$ given by eq.\ (\ref{uinitial}).
The choice $\bar{\la}_{1\La}/\La=0.1$, 
$\bar{\la}_{2\La}/\La=2$ corresponds to a strong first order
phase transition with renormalized masses
not much smaller than the cutoff 
scale $\La$. In this case the $k$-dependence of the 
effective average potential resembles the one 
presented in fig.\ 1.
In contrast
we give two examples where the dependence of 
$\si_k$ on the coarse graining scale becomes of crucial
importance.
The choice $\bar{\la}_{1\La}/\La=2$, 
$\bar{\la}_{2\La}/\La=0.1$ leads to a weak first
order phase transition with small renormalized masses
and the behavior of the effective average potential
is similar to the one given in fig.\ 3. 
The coarse grained potential 
and the surface tension show 
a high sensitivity on the scale $k$.
A more increased sensitivity on the scale $k$
can be observed
for $\bar{\la}_{1\La}/\La=4$,
$\bar{\la}_{2\La}/\La=70$ which corresponds to a 
relatively strong
first order phase transition. 

The $k$-dependence of the surface tension $\si_k$ is displayed in
fig.\ 5. Here $\si_k$ is normalized to 
its maximum value $\si_*$ where
$\si_*/\La^2=1.67 \times 10^{-2} 
(8.41 \times 10^{-11})
(1.01 \times 10^{-3})$ for 
$\bar{\la}_{1\La}/\La=0.1(2)(4)$, 
$\bar{\la}_{2\La}/\La=2(0.1)(70)$. 
The scale $k_f$ is given by $|m_{max}|$
with $m_{max}^2=
2 Z_k^{-1} \rho_{max} (\prl^2 U_k/\prl \rho^2)(\rho_{max})$
denoting the renormalized
mass term at the top of the potential barrier
$\rho_{max}$. More precisely, we chose
$(k_f^2-|m_{max}^2(k_f)|)/k_f^2=0.01$.
For $\bar{\la}_{1\La}/\La=0.1$,
$\bar{\la}_{2\La}/\La=2$ the curve exhibits a small 
$k$-dependence
around its maximum and $\si_* \simeq \si_{k_f}$.
For the second and the third example 
one observes that
$\si_{k_f}$ becomes considerably smaller than the maximum
value due to a strong $k$-dependence.
The coarse graining scale $k$ should not be taken smaller
than the inverse bubble wall thickness
$\Dt^{-1}$ \cite{matrix}. This ensures that the
detailed properties of the bubble are irrelevant
for the computation of the `classical' 
surface tension $\si_k$. 
We estimate the bubble wall thickness $\Dt$ by
\be
\Dt= 2 Z_k \frac{\rho_{min}}{\si_k}
\ee
where we have 
taken the gradient energy as half the total 
surface energy and 
approximated the mean field gradient
at the bubble wall by $\vp_{min}/\Dt$.
For the given examples we observe
$\Dt^{-1}(k_f) \simeq k_f/4$. We choose
$k \simeq k_f$ as the coarse graining scale. 

In order to quantify the differences between the three 
examples we have displayed some characteristic 
quantities in the table.
The renormalized couplings
\be
\la_{1R}=Z_{k_f}^{-2} 
\frac{\prl^2 U_{k_f}}{\prl \rho^2}(\rho_{min}),\qquad
\la_{2R}=4 Z_{k_f}^{-2} 
\frac{\prl U_{k_f}}{\prl \tau}(\rho_{min})
\ee
are normalized with respect to the mass term
\be
m_{R}^c=(2 Z_k \rho_{min} \la_{1R})^{1/2} \, .
\ee
In addition we give
the mass term
\be
m_{2R}^c=(Z_k \rho_{min} \la_{2R})^{1/2}
\ee
corresponding to the curvature of the potential in the direction
of the second invariant $\tau$. 
All couplings and masses are evaluated at the 
critical temperature (critical $\rho_{0\La}$). 
In comparison with 
fig.\ 5 one observes in the vicinity of $k_f$ a weaker scale 
dependence of $\sigma_k$
for smaller effective couplings. In particular, a
reasonably weak scale dependence of $U_k$ and $\si_k$ requires
\be
\frac{\la_{1R}}{m_R^c}
\ll 1 \label{weakk}\, \, . 
\ee
This establishes a quantitative criterion for the range
where Langer's theory can be used without paying too much
attention to the precise definition of the coarse graining.
In the table we also present $k_f/\La$ and the renormalized 
masses in units of $\La$, 
which indicate the strength of the phase transition.
In particular,
for the relatively strong phase transition
of the third example
with slightly larger effective couplings 
one observes an increased scale 
dependence as compared to the weak phase transition
of the second example. 
This clearly shows that the `strength' of the phase
transition is, in general, not the primary criterion
for the applicability of Langer's theory.

In addition to the dependence on $k$, the coarse grained 
free energy depends also on the precise shape
of the infrared cutoff function or the
inverse average propagator $P_k(q)$ given by eq.\ (\ref{two}). 
Analytical studies in the Abelian Higgs model 
indicate \cite{litim} that this scheme dependence
is rather weak  
for the effective potential and the
surface tension. 

In summary, we have shown that 
the coarse grained free energy cannot be defined
without detailed information on the coarse graining scale
$k$ unless the effective dimensionless couplings are small
at the phase transition. For Abelian and non-Abelian gauge theories
with a small gauge coupling at the scale $k=\La$ this coincides
with a relatively strong first order transition.
Only for small
couplings a range with a weak $k$-dependence 
of the classical surface tension appears. There is
a close relation between the dependence of the coarse grained
free energy on the coarse graining scale and the reliability
of the saddle point approximation in Langer's theory
of bubble nucleation. For a strong $k$-dependence of $U_k$
a small variation in the coarse graining scale can induce 
large changes in the predicted nucleation rate
in lowest order in a saddle point approximation.
In this case the $k$-dependence of the pre\-factor
$A_k$ has also to be computed. Therefore, for strong dimensionless
couplings a realistic estimate of the nucleation rate
needs the capability to compute
$\ln A_k$ with the same accuracy as $16 \pi \sigma_k^3/3 \eps^2$
and a check of the cancelation of the $k$-dependence
in the combined expression (\ref{nc}). 
Our observation that the details of the coarse graining 
prescription become less important in the case of small
dimensionless couplings is consistent with the
fact that typically small couplings are needed for a
reliable saddle point approximation for $A_k$.

\newpage

\section*{Figures}

\begin{enumerate}

\item  
Evolution of the potential for a strongly first order
phase transition. 
\vspace{5mm}

\item  
Characteristics of the potential for a strongly first order
phase transition in dependence on the coarse graining scale $k$.
\vspace{5mm}

\item  
Evolution of the potential for a weakly first order
phase transition. 
\vspace{5mm}

\item  
Characteristics of the potential for a weakly first order
phase transition in dependence on the coarse graining scale $k$.
\vspace{5mm}

\item  
The normalized surface tension
$\si_k/\si_*$ as a function of $\ln (k/k_f)$.
The short distance parameters are (1)
$\bar{\la}_{1\La}/\La=0.1$, $\bar{\la}_{2\La}/\La=2$,
(2) $\bar{\la}_{1\La}/\La=2$, $\bar{\la}_{2\La}/\La=0.1$,
(3) $\bar{\la}_{1\La}/\La=4$, $\bar{\la}_{2\La}/\La=70$. 
\vspace{5mm}

\end{enumerate}

\section*{Table}

\begin{enumerate}

\item  
Effective dimensionless renormalized couplings
$\la_{1R}/m_R^c$ and $\la_{2R}/m_R^c$
at the phase transition. The critical couplings 
and mass
terms $m_R^c$, $m_{2R}^c$ are evaluated at the scale $k_f$.
\label{table2}
\vspace{5mm}

\end{enumerate}

\begin{figure}
\unitlength1.0cm
\begin{center}
\epsfysize=15.cm
\epsfxsize=15.cm
\epsffile{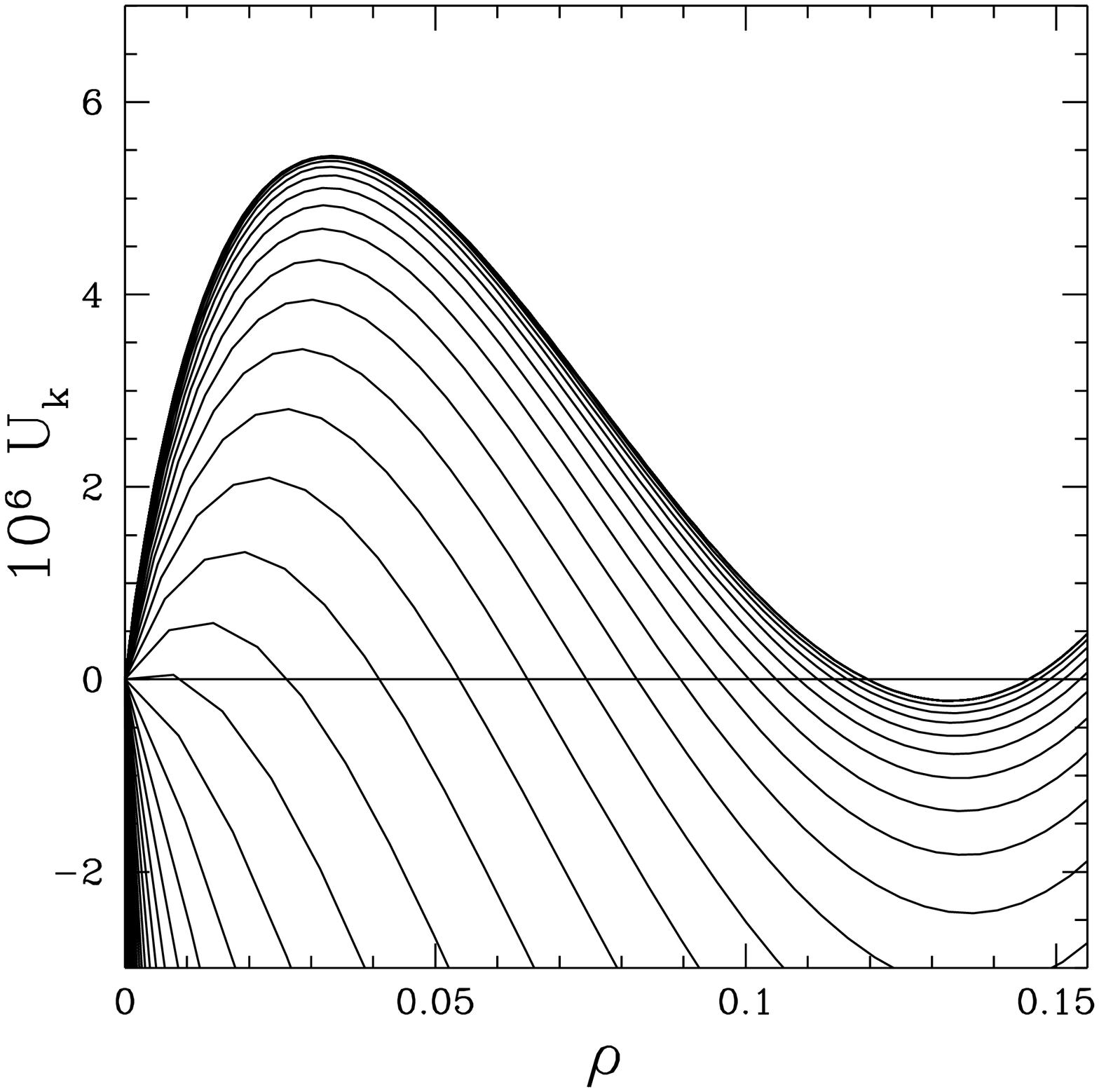}
\end{center}
\end{figure}  

\begin{figure}
\unitlength1.0cm
\begin{center}
\epsfysize=15.cm
\epsfxsize=15.cm
\epsffile{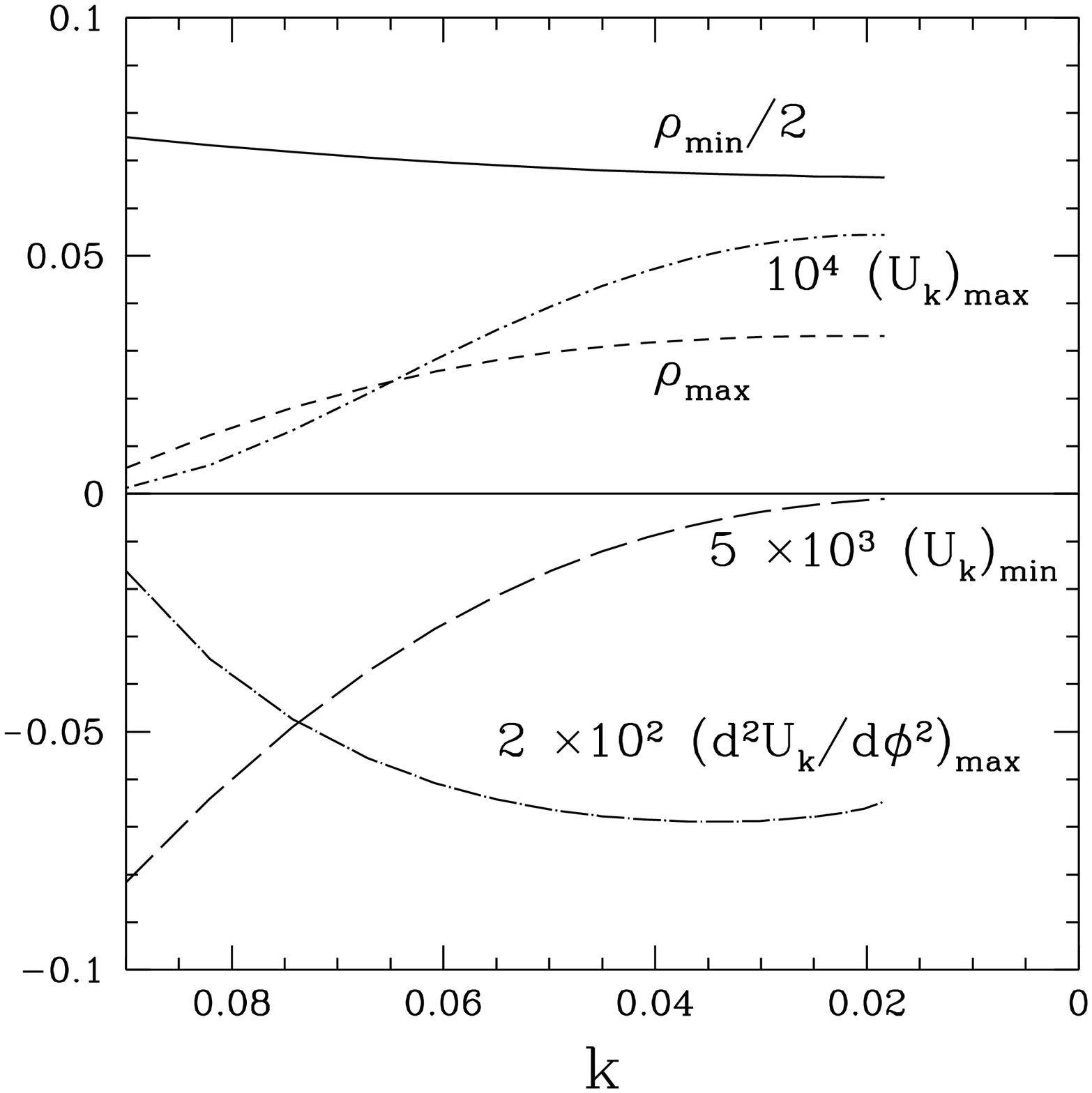}
\end{center}
\end{figure}  

\begin{figure}
\unitlength1.0cm
\begin{center}
\epsfysize=15.cm
\epsfxsize=15.cm
\epsffile{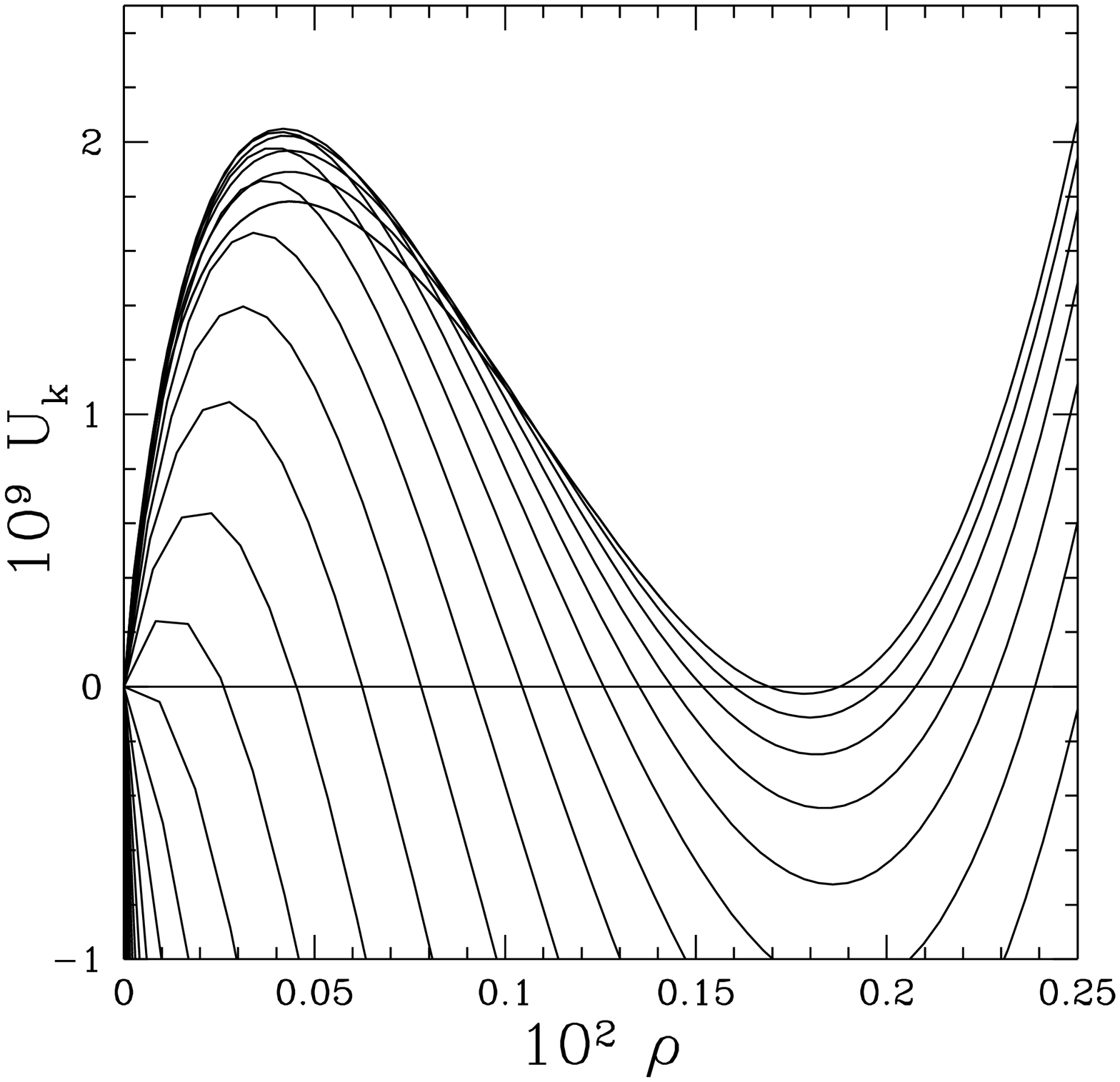}
\end{center}
\end{figure}  

\begin{figure}
\unitlength1.0cm
\begin{center}
\epsfysize=15.cm
\epsfxsize=15.cm
\epsffile{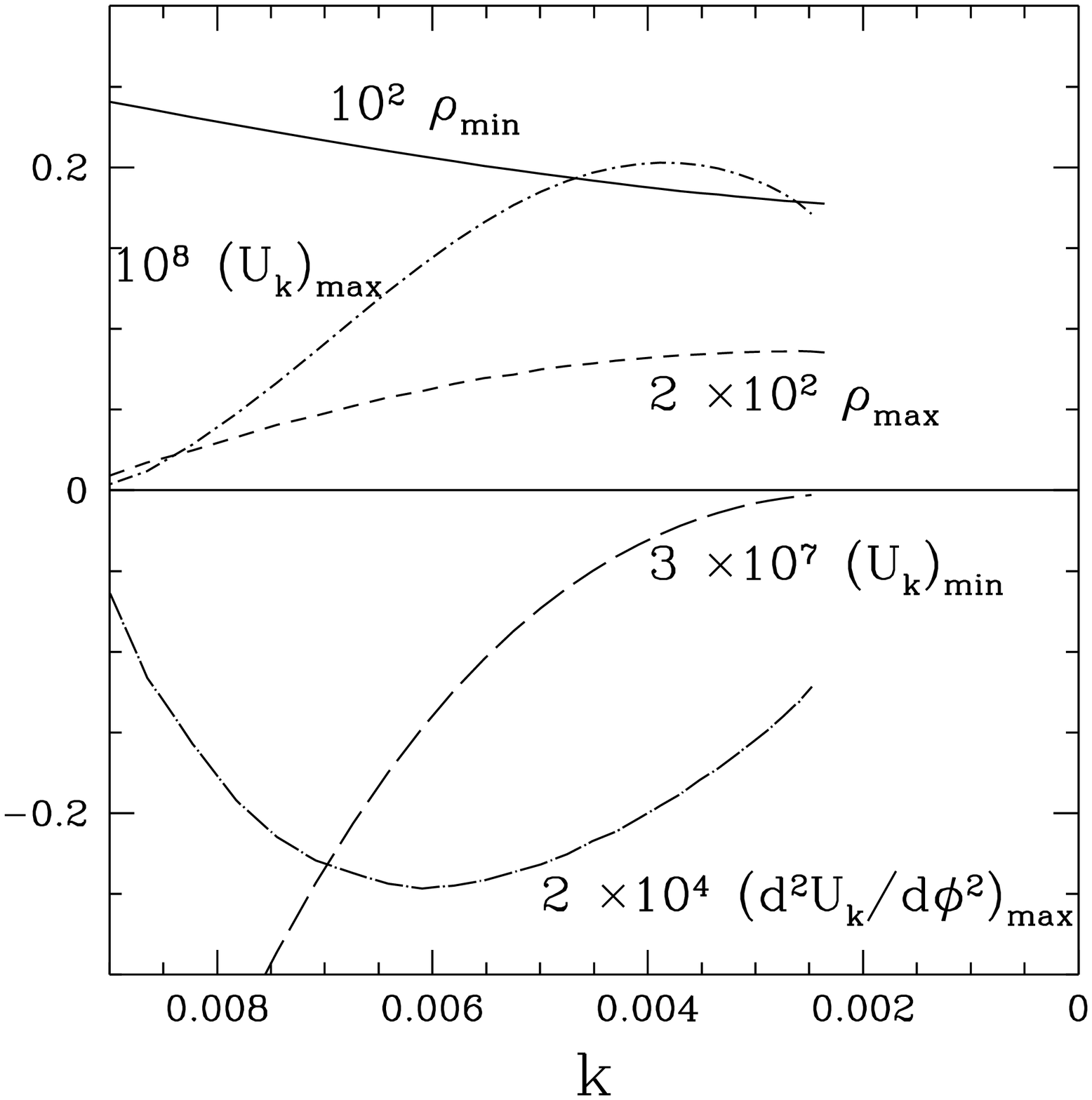}
\end{center}
\end{figure}

\begin{figure}
\unitlength1.0cm
\begin{center}
\begin{picture}(13.,9.)
\put(-0.7,4.6){\large $\ds{\frac{\si_k}{\si_{*}}}$}
\put(6.,-1.){\large $\ds{\ln\left({k}/{k_f}\right)}$}
\put(6.9,2.){(1)}
\put(9.2,2.){(2)}
\put(11.3,2.){(3)}
\put(-0.4,0.){
\epsfysize=13.cm
\epsfxsize=9.cm
\rotate[r]{\epsffile{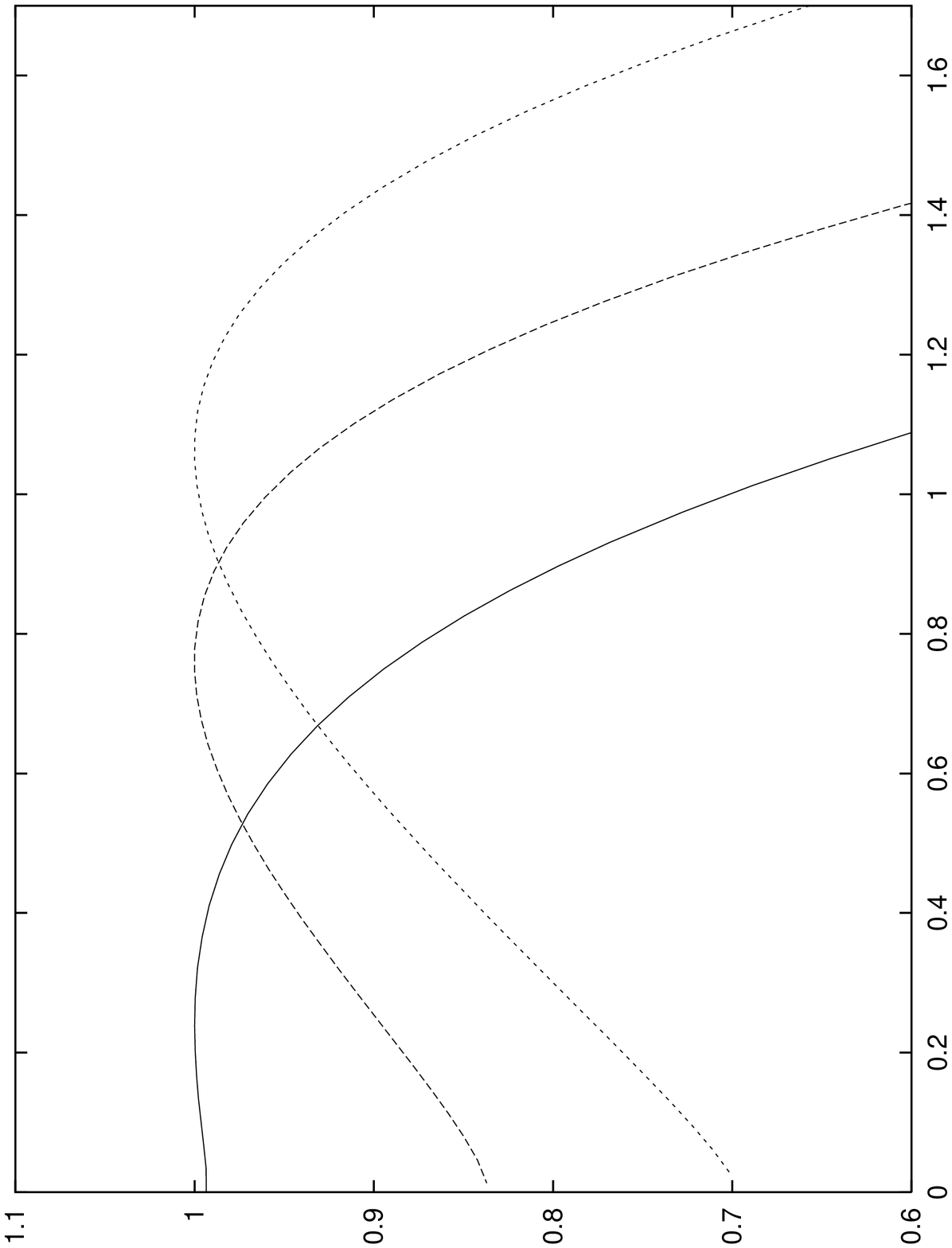}}
}
\end{picture}
\end{center}
\end{figure}

\begin{table} [h]
\renewcommand{\arraystretch}{2.0}
\hspace*{\fill}
\begin{tabular}{|c|c|c|c|c|c|c|c|}     \hline

$\ds{\frac{\bar{\la}_{1\La}}{\La}}$
&$\ds{\frac{\bar{\la}_{2\La}}{\La}}$
&$\ds{\frac{\la_{1R}}{m_R^c}}$
&$\ds{\frac{\la_{2R}}{m_R^c}}$
&$\ds{\frac{m_R^c}{m_{2R}^c}}$
&$\ds{\frac{m_R^c}{\La}}$
&$\ds{\frac{m_{2R}^c}{\La}}$
&$\ds{\frac{k_f}{\La}}$
\\ \hline \hline
0.1
& 2
&0.228
&8.26
&0.235
&$1.55 \times 10^{-1}$
&$6.62 \times 10^{-1}$
&$1.011 \times 10^{-1}$
\\ \hline 
$2$
&0.1
&0.845
&15.0
&0.335
&$2.04 \times 10^{-5}$
&$6.10 \times 10^{-5}$
&$1.145 \times 10^{-5}$
\\ \hline 
4
&70
&0.980
&16.8
&0.341
&$6.96 \times 10^{-2}$
&$2.04 \times 10^{-1}$
&$3.781 \times 10^{-2}$
\\ \hline 
\end{tabular}
\end{table}

\end{document}